\documentclass{sig-alternate-10pt}
\usepackage[margin=1in]{geometry}

\usepackage[margin=1in]{geometry}

\usepackage{booktabs} 

\usepackage[linesnumbered,vlined, boxed, ruled]{algorithm2e}

\SetAlFnt{\small}
\SetAlCapFnt{\small}
\SetAlCapNameFnt{\small}
\SetAlCapHSkip{0pt}
\IncMargin{-\parindent}

\usepackage{amsthm}
\usepackage{xspace}
\usepackage{graphicx}
\usepackage{subfigure}
\usepackage{mathtools}
\usepackage{color}

\usepackage[linesnumbered,vlined, boxed, ruled]{algorithm2e}
\usepackage{epsfig}
\usepackage{epstopdf, url}

\usepackage[T1]{fontenc}
\usepackage[utf8]{inputenc}
\usepackage[font=small,labelfont=bf]{caption}
\usepackage{booktabs}

\SetKwInOut{Input}{input}\SetKwInOut{Output}{output}
\SetKwComment{Comment}{//}{}
\SetEndCharOfAlgoLine{}
\SetKw{True}{true}
\SetKw{False}{false}
\SetKw{Not}{not}

\newcommand{\rev}[1]{{{\bf }#1}}

\newcommand{\eat}[1]{}

\begin{document}
	\pagestyle{empty}
	\title{The World of Graph Databases from An Industry Perspective \vspace*{-5mm}}
	
\author{
	\alignauthor Yuanyuan Tian\\
	\affaddr{Gray Systems Lab, Microsoft}\\
	\email{yuanyuantian@microsoft.com}
}
\maketitle
	\begin{abstract}
	Rapidly growing social networks and other graph data have created a high demand for graph technologies in the market. A plethora of graph databases, systems, and solutions have emerged, as a result. On the other hand, graph has long been a well studied area in the database research community. Despite the numerous surveys on various graph research topics, there is a lack of survey on graph technologies from an industry perspective. The purpose of this paper is to provide the research community with an industrial perspective on the graph database landscape, so that graph researcher can better understand the industry trend and the challenges that the industry is facing, and work on solutions to help address these problems. 
	\end{abstract}
	
	\section{Introduction}
	
	Rapidly growing social networks and other graph data have created a high demand for graph technologies. No wonder Gartner ranked graph technologies among the top 10 data and analytics trends in 2021~\cite{gartner}. According to Gartner, up to 50\% of their client inquiries around the topic of AI involve a discussion about the use of graph technology~\cite{gartner}, and by 2025,  graph  technologies  will  be  used  in  80\%  of  data  and  analytics  innovations~\cite{gartnerGraphGuide}.  Inkwood Research projected that the global market for graph databases will grow at  21.7\% from 2019 to 2027, and reach \$4.6 billion by 2027~\cite{inkwood}. The industry has responded to the high demand of graph technologies with a boom of graph companies, systems, and solutions, as depicted in~\cite{graphlandscape}. The venture capital investment has also been very active in graphs in the last couple of years. Not only new startups, like Katana graph (\$28.5 million in Series A), but even seasoned graph database companies, like Neo4j and TigerGraph, received a lot of funding (Neo4j raised \$325 million in Series F and TigerGraph received \$105 million in Series C).
	
	
	On the research side, graph has long been a well studied area in the database research community. 
	In his VLDB 2019 keynote~\cite{graphkeynote}, Professor Özsu provided a good summary of the various subareas of graph research. \rev{Professor Boncz delivered a keynote in EDBT 2022 about the state of graph database systems \cite{edbtkeynote}, touching on graph models, graph languages, the common pitfalls in designing graph database systems, and the blueprint of a competent graph database system. Professor Fan's keynote in VLDB 2022~\cite{fankeynote} discussed the challenges and progress made on processing big graphs, including parallel scalability, incremental computation, and semantic joins between relations and graphs. There have also been numerous research surveys on topics such as graph database models~\cite{graphdbmodel}, graph query languages~\cite{graphLanguages}, graph stream algorithms~\cite{graphstream}, knowledge graphs~\cite{kg}, distributed graph pattern matching~\cite{patternmatch}, large-scale graph processing~\cite{mybook}, etc. Back in 2014, Professor Deshpande blogged his views on graph data management and pointed out some open problems~\cite{blog}. The VLDB 2018 best paper~\cite{semihsurvey} and its extension~\cite{surveyjournal} conducted a comprehensive user survey about how graphs are used in practice, and revealed many interesting insights, including the ubiquity of large graphs, variety of entities represented by graphs, the scalability challenges faced by many graph systems, the importance of visualization tools, and the continued popularity of RDBMSs in managing and processing graphs. The recent community publication~\cite{futureview} painted a picture of what the next-decade big-graph processing systems look like in the aspects of abstractions, ecosystems, and performance. However, none of the above work discussed in detail the solution space or architecture of existing graph databases in the market. Despite the recent surge in graph technology innovation in the industry, there is still a lack of survey on graph technologies from an industry perspective.}

	The database research community, as a whole, has been having very strong ties to and impact on the industry, witnessed by the fleet of database products (e.g. PostgreSQL and Flink) and startups (e.g. Vertica and Databricks) originated from research. In the area of graph databases, the research community has also influenced heavily on graph benchmarking~\cite{ldbcbenchmark} and graph query languages~\cite{gql}. But still, some of the major problems that the graph database industry cares about are not well known to the research community. The purpose of this paper is to provide the research community with an industrial perspective on the graph database landscape, in the hope of helping researchers better understand the current industry status quo and the challenges they are facing, and ultimately increasing the impact of the graph database research community.

\section{Use Cases  and Workloads}\label{sec:use-case}
	
	
In terms of customer use cases, graph databases have been used in many vertical industries, including finance, insurance, healthcare, retail, energy, power, manufacturing, government, marketing, supply chain, transportation, etc. \rev{This diverse and wide applicability of graphs in many domains is also observed in~\cite{semihsurvey}.} Some of the concrete use cases of graph databases have been provided in~\cite{neo4jUsecase, oraclebook, tigergraphusecases, db2graph}. Perhaps, the most common example of graph database usage is fraud detection. For example, \cite{db2graphdemo} demonstrated a detailed example scenario of traversing through a graph containing insurance claims information and patients medical records to detect fraudulent claims.

Similar to the different types of workloads in relational databases, there are also two different types of graph database workloads. The first type focuses on low-latency graph traversal and pattern matching. They are often called \emph{graph queries}. These queries only touch small local regions of a graph, for example, finding 2-hop neighbors of a vertex, or the shortest path between two vertices. Due to the low-latency requirement and the interactive nature of the graph queries, people also call them \emph{graph OLTP}. Graph OLTP is often used in exploratory analysis and case studies. The second type of graph workload is \emph{graph algorithms}, which usually involve iterative, long running processing on the entire graph. Good examples are Pagerank and community detection algorithms. Graph algorithms are often used for BI-ish applications. Because of this reason, people also call them \emph{graph OLAP}. Recently, a new trend emerges that combines graph and machine learning together, called \emph{graph ML}. For example, graph embedding or vertex embedding are used to transform graph structures into vector space which are then included as features for ML model training. Graph neural network (GNN) is another example of graph ML. Quite often graph ML is lumped together with the graph OLAP workload.


\section{Graph Models}\label{sec:model}

\begin{figure}[tbh]
	\centering
	\subfigure[RDF Model]{
		\label{fig:rdf}\includegraphics[width=0.7\linewidth]{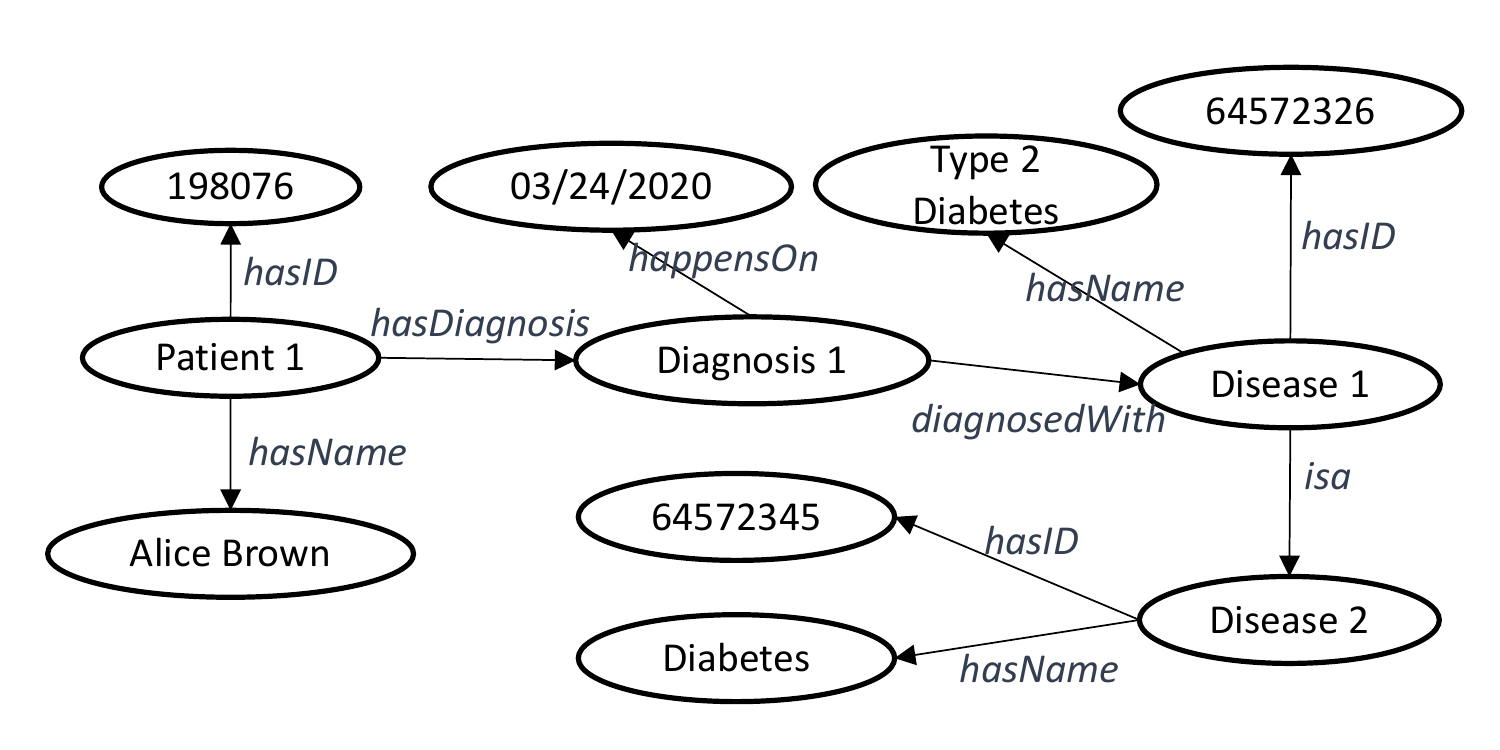}} 
	\subfigure[Property Graph Model]{
		\label{fig:pg}\includegraphics[width=0.7\linewidth]{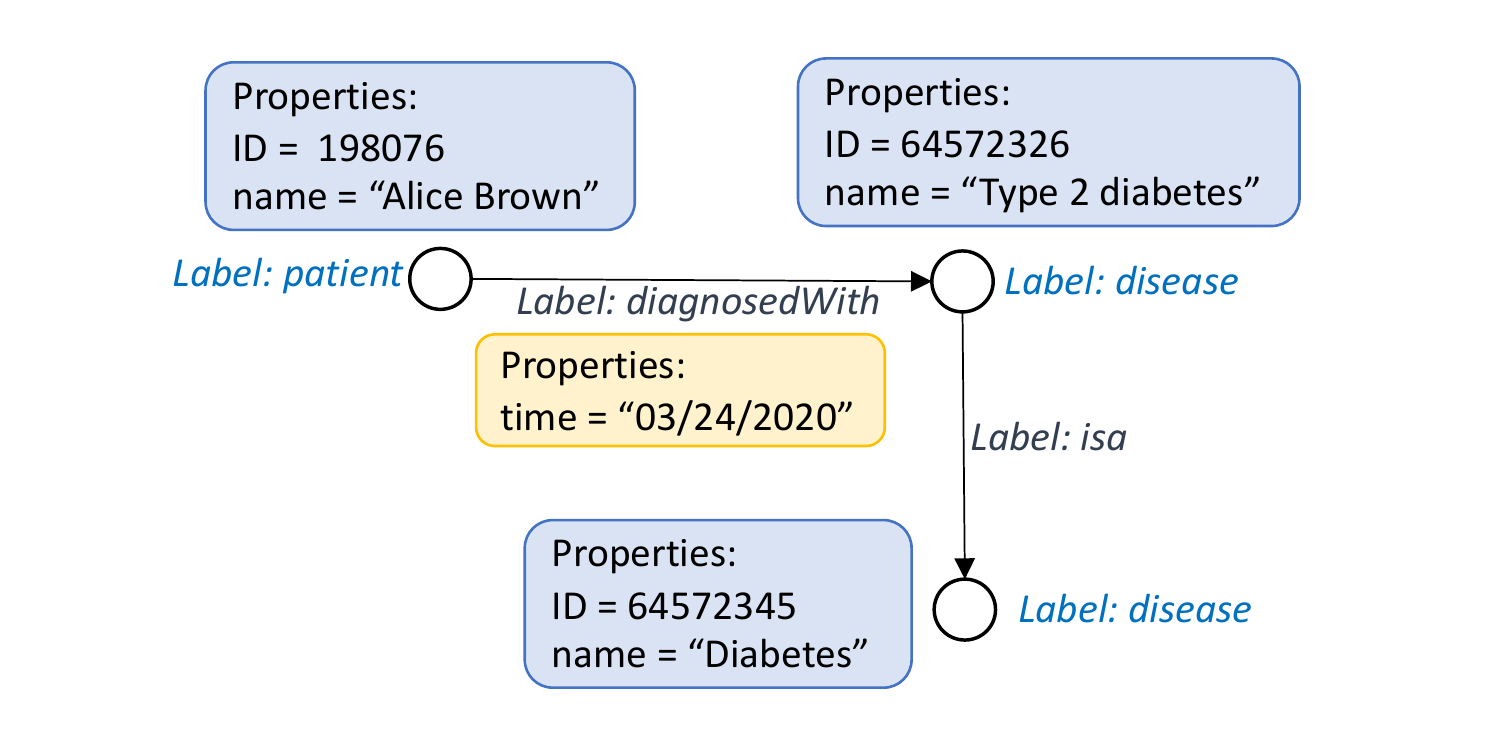}} 
	\vspace*{-2mm}
	\caption{\label{fig:model} RDF and property graph models}
		\vspace*{-2mm}
\end{figure}

Whenever talking about a graph database, we need to first talk about the graph model(s) that it supports. The two prominent graph models supported by most commercial graph databases are the RDF model and the property graph model. 

\textbf{RDF Model.}
RDF is among the suite of W3C standards to support Linked Data and Knowledge Graphs~\cite{rdf}. An RDF graph is a directed edge-labeled graph, represented by the subject–predicate–object triples. Figure~\ref{fig:rdf} shows an example graph represented in the RDF model. This graph captures the following information: A patient, named Alice Brown, with patient ID 19806, is diagnosed with Type 2 Diabetes which has disease ID 64572326 on March 24, 2020; and Type 2 Diabetes is sub-type of Diabetes which has disease ID 6472345. For example, in the (Patient 1) $-$[hasName]$\rightarrow$ (Alice Brown) triple, Patient 1 is the subject, hasName is the predicate, and Alice Brown is the object. The RDF model is often used in knowledge representation and inference as well as sematic web applications. \rev{For example, DBPedia~\cite{dbpedia} and YAGO~\cite{yago} both utilize RDF to represent their knowledge graphs and support queries on the knowledge bases using SPARQL~\cite{sparql}.}

\textbf{Property Graph Model.}
In comparison, a property graph is a direct graph where each vertex and edge can have arbitrary number of \emph{properties}. Vertices/edges can also be tagged with \emph{labels} to distinguish the different types of objects/relationships in the graph. Figure~\ref{fig:pg} shows how the same information captured in the RDF graph in Figure~\ref{fig:rdf} is represented in the property graph model. Here, instead of representing the ID and the name of a patient or disease as separate nodes, the property graph model can fold them in as the properties of the patient and the disease nodes. Similarly, the diagnosis time can be represented as a property of the diagnosedWith edge, eliminating the need to create a separate diagnosis node and its connecting edges to the patient and disease nodes. In general, the property graph model can capture the same information with fewer nodes and edges than the RDF model, \rev{as illustrated by this example.} \rev{This is because a piece of information can only be represented either as a node or an edge in the RDF model, whereas the property graph model can also define it as an attribute of an existing node or edge, thus leading to fewer number of nodes and edges in the graph.} The property graph model is often used for applications that require graph traversal, pattern matching, path and graph analysis. 

Today, although both models are supported in the graph database industry, \rev{as we will show in Section~\ref{sec:offerings}}, the property graph model has overwhelming endorsement, despite the fact that RDF is a much older model. \rev{All the major offerings we surveyed in the paper support the property graph model, and two of them also support the RDF model. } In~\cite{rdfpg}, Hartig proposed a formal transformations between the RDF and property graph models, in the hope to reconcile both models.

	
\section{Graph Query Languages}\label{sec:language}


On the graph OLTP side, for RDF graphs, there is the standard SPARQL query language~\cite{sparql}. For property graphs, there are many languages being used and proposed, but no clear winner. One of the top contenders is Tinkerpop Gremlin~\cite{tinkerpop} which is supported by around 30 graph vendors today, also probably the most widely used graph query language today. 
Another strong contender is openCypher~\cite{opencypher}. Cypher~\cite{cypher} was originally Neo4j’s proprietary declarative graph query language, and it was open-sourced in 2015.  Around 10 graph vendors support openCypher. 
Besides these two more widely adopted languages, many vendors proposed their own graph query languages. Oracle proposed a declarative language based on SQL, called PGQL~\cite{pgql}. GSQL~\cite{gsql} is the SQL-like graph query language adopted by TigerGraph. Microsoft SQL Graph extended SQL with MATCH clause for graph pattern matching~\cite{sqlgraph}. The LDBC~\cite{ldbc} Graph Query Language Task Force (with members from both academia and industry) has proposed G-Core~\cite{gcore}. In an attempt to reduce the chaos on graph query languages, in 2019, the Joint Technical Committee 1 of ISO/IEC, approved a project to create a standard graph query language, called GQL~\cite{gql}. This effort is also complemented by another project that extends SQL with graph view definition and graph query constructs, called SQL/\rev{PGQ}. GQL and SQL/PGQ share a common declarative graph pattern matching language component. This common component integrates ideas from openCypher, Oracle’s PGQL, TigerGraphs’s GSQL, and LDBC G-CORE. The standardization effort on GQL and SQL/\rev{PGQ} has strong participation from academia, and it is one of the areas where the graph research community has heavily impacted the graph industry. 
\rev{However, given the current state of the graph query languages, even when GQL and SQL/PGQ standards are published, it is going to take time for the vendors to adopt it, since a large number of graph applications are already written in these existing languages.}
It will still take many years for the standardization to settle down.


In terms of language properties, Gremlin is more of an imperative graph traversal language (although the recent version of Gremlin also has some declarative language features), while the others are declarative. As a result, Gremlin is relatively more low-level and less user-friendly. But in terms of expressiveness, Gremlin is Turing Complete~\cite{gremlinTuringComplete}, while most of the declarative counterparts, including openCypher, are not. This means there are graph algorithms or operations not expressible in these non-Turing-Complete languages. Out of all the declarative languages, TigerGraph's GSQL is the only one that is Turing Complete~\cite{qsqltc}.  


On graph OLAP side, there is also no standard language or API, but most vendors support a variation of the Pregel-like API~\cite{pregel}. Like for ML, a library of build-in graph algorithms are more useful to users, so the lack of standard is not so much an issue for graph OLAP.

	
\section{Graph Database Offerings}\label{sec:offerings}

\begin{figure*}
		\includegraphics[width=0.9\linewidth]{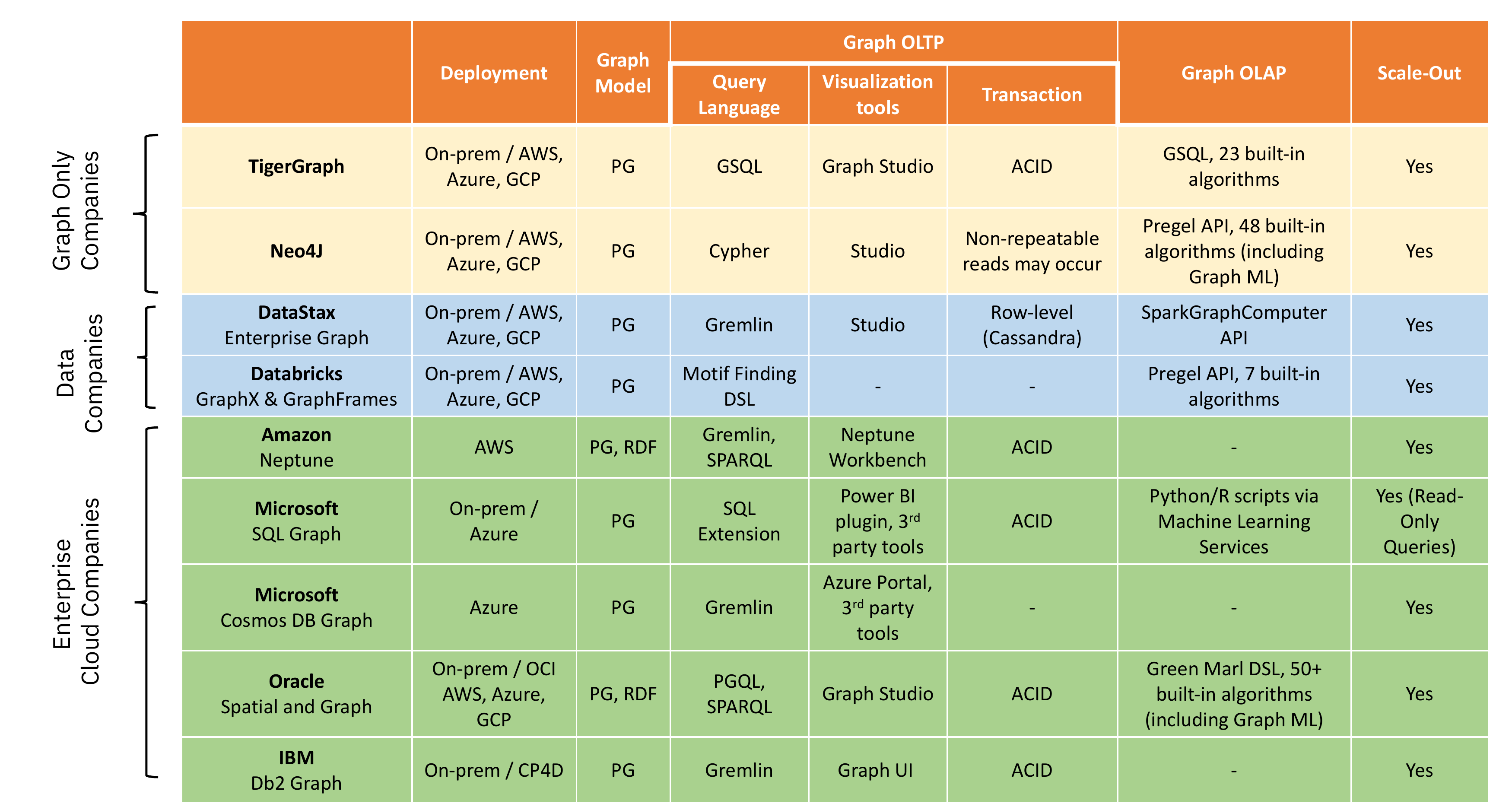}
		\caption{\label{fig:offerings} Major Graph Database Offerings}
		\vspace*{-2mm}
	\end{figure*}

The graph database area is a very crowded space in the industry, with new projects and startups popping out every moment. It is impossible to enumerate all the current graph database offerings.  So, this section only highlights some of the major offerings in the three categories: graph-database-only vendors, \rev{data companies} with graph support, and enterprise cloud vendors with built-in graph database support. The different features of their graph products are summarized in Figure~\ref{fig:offerings}. In Section~\ref{sec:solution}, we will discuss the different architecture solutions adopted by the various vendors.

In the pure-play space, Neo4j and TigerGraph are the two strongest contenders. They provide solutions both on premise and on all major clouds, AWS, Azure, and GCP. They have good support for both graph OLTP and OLAP workloads. The pure players have also perfected the art of visualization and tooling, as well as the support of a large number of built-in graph algorithms.

DataStax and Databricks are two data companies with a wide range of data portfolio. The graph component is also integrated well with the rest of the system components. For example, DataStax Enterprise Graph (DSG) is built on top of the DataStax's main NoSQL data engine Cassandra. And for Databricks graph support, GraphX is built on top of Spark's RDD, and GraphFrames is based on DataFrames. Since both companies aim at more general data systems, their support on graphs is not as comprehensive as the graph-only vendors. DataStax's support on graph OLAP is very rudimentary \rev{(only relying on SparkGraphComputer API in Gremlin with just 3 built-in graph algorithms)}. Databricks' graph OLTP support comes only from the simple motif finding support in GraphFrames. This support is not only limited by the very simple motif finding DSL, but also unlikely to perform well, since the graph OLTP query processing utilizes DataFrames underneath, which is originally designed for analytics purpose. 

The last category of graph database vendors are big cloud companies, including Amazon, Microsoft, Oracle, and IBM. They all provide a large number of data services on their cloud platform, and the built-in graph database service is one of them. Microsoft, Oracle, and IBM are previously big relational database shops, so it is not surprising that their graph database solutions are based on their relational databases: Microsoft SQL Graph on top of SQL Server (on premise) and Azure SQL Database (on cloud), Oracle Spatial and Graph on top of Oracle databases (on premise and cloud), and IBM Db2 Graph on top of Db2 (on premise and Cloud Pak for Data). Microsoft also provides another graph database solution, Cosmos DB Graph, built on top of the NoSQL database Azure Cosmos DB. Amazon, on the other hand, builds Neptune on the same back end storage as other AWS platforms, such as Aurora and DynamoDB. Amazon, since a pure cloud company, doesn't provide an on-premise graph database solution. Most of the graph databases from this category focus on the graph OLTP workload, except that Oracle Spatial and Graph has a very good graph OLAP support with a large number of built-in algorithms. 

Now, let's look at the different dimensions of the table shown in Figure~\ref{fig:offerings}. Most vendors support both on-premise and cloud deployment, except that Amazon Neptune and Microsoft Cosmos DB Graph are on cloud only. In terms of graph models, all the vendors support the property graph model. Amazon Neptune along with Oracle Spatial and Graph additionally support the RDF model. For graph OLTP workload, the languages that different vendors use reflect the language chaos discussed in Section~\ref{sec:language}, but Gremlin appears to be the most supported. Due to the exploratory nature of graph OLTP workload, visualization is especially important for customers. \rev{Most graph vendors do provide visualization support. Compared to relational databases, transaction support has been a sore spot for graph databases. An update to a single node often affected its edges and connected nodes, e.g. deleting a node requires the deletion of all the edges it is connected to. So, a transaction is often more complex in a graph database, especially in a distributed setting. Some graph databases manage to provide full ACID support, but others either have no support or week support for transactions.} Compared to graph OLTP, the graph OLAP support is relatively weak in general, but TigerGraph, Neo4j, and Oracle stand out due to their large number of built-in algorithms. \rev{The VLDB survey~\cite{semihsurvey} has observed the ubiquity of large graphs with over a billion edges and pointed out that scalability is a challenge that many users face. As a result, major graph vendors strive to address this challenge.}  All the graph database solutions can scale up nicely to a certain extent, which can satisfy a lot of customers, and most also provide scale-out solutions for those customers working on huge graphs that cannot fit in a single node. \rev{As rightly pointed out by~\cite{fankeynote}, while distributed parallelization can handle larger graphs, it does not always provide desirable performance. Due to the connected nature of graphs, it is almost impossible to achieve access locality in a distributed setting. As a result, distributed graph computation often accesses many partitions of a graph, which incurs a lot of communication cost. If a large graph can fit in a single node, the scale-up solution might provide better performance than the scale-out version of the same system. As shown in~\cite{db2graph}, a single node system on a decent machine configuration can comfortably handle large graphs with billions of edges. However, concurring with~\cite{semihsurvey}, efficiently querying and processing large-scale graphs (way beyond billions of edges) remains a challenge. }

		
\section{Graph Database Solution Space}\label{sec:solution}
	
	\begin{figure}
		\centering
		\includegraphics[width=0.9\linewidth]{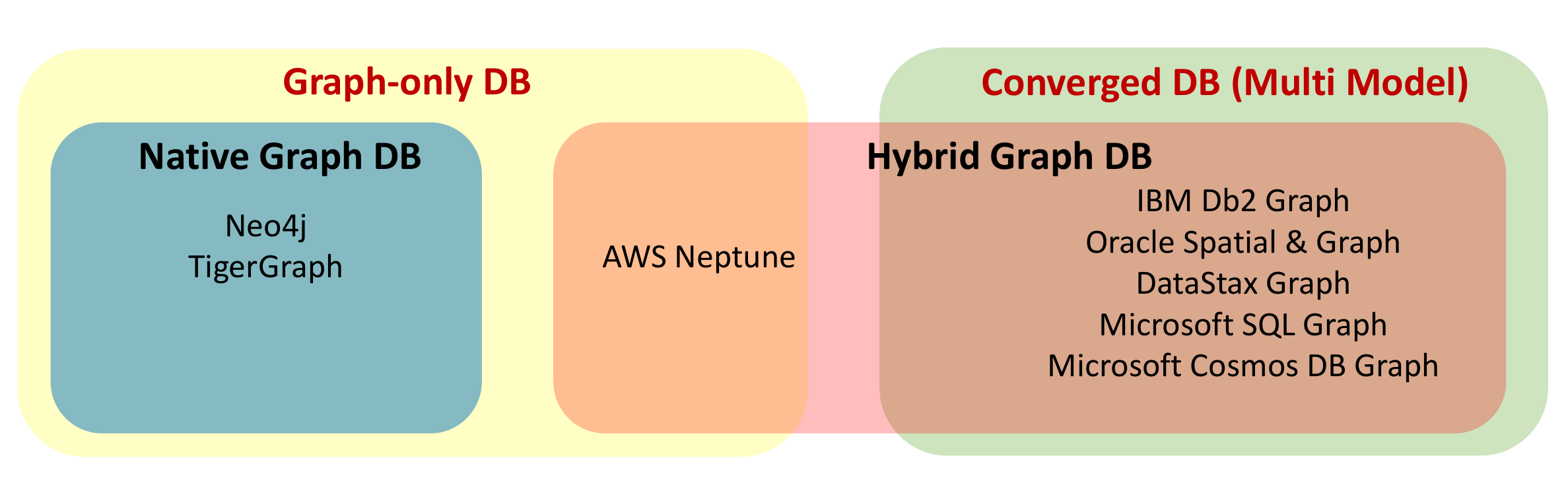}
		\caption{\label{fig:solutions} Graph Solution Space}
		\vspace*{-2mm}
	\end{figure}
	
	
\subsection{Native vs Hybrid Graph Databases}
	
One way to categorize the solution space is \textit{native} graph databases vs \textit{hybrid} graph databases, as shown in Figure~\ref{fig:solutions}. As the name suggested, native graph databases are built with specialized query and storage engines from scratch just for graphs. Neo4j and Tigergraph are two prime examples of native graph databases. \rev{This type of graph databases are highly optimized to the supported graph workloads.} But the drawback is the high engineering cost, since they have to reinvent the wheels for the support of transactions, access control, scalability, high availability (HA), disaster recovery (DR) and so on. In contrast, a hybrid graph database has a specialized graph query engine, but resorts to an existing data store to handle the persistence of data, be it either a SQL database, a key value store, or a document store. As shown in Figure~\ref{fig:solutions}, more graph databases fall into this camp. Since a hybrid graph database delegates its storage engine to an existing data store, it has much faster development time. In addition, it can also get many things for free from the backend store, such as transaction support, access control, scalability, HA and DR, etc. But the potential downside is that the performance of a hybrid graph database may not match a highly optimized native graph database. Of course, the performance of an individual graph database also highly depends on the implementation details.
	
\subsection{Graph-Only vs Converged Databases}

Another way to categorize the solution space is \textit{graph-onl}y databases vs \textit{converged} databases, or also called \textit{multi-model} databases. As shown in Figure~\ref{fig:solutions}, all native graph databases are graph-only databases, and most hybrid graph databases are converged databases, but some are graph-only. Graph-only databases support graph workload only. This fact can also be a fundamental limitation of these databases. In fact, Neo4j and TigerGraph both have dedicated chapters on data import and export in their user manuals. In contrast, a converged database or a multi-model database supports poly query languages/APIs on the shared data. This fact can also be an advantage of the converged database architecture. We elaborate on some of the advantages below.

	Fundementally, the converged database solution solves the \textit{fragmented} database problem. Real applications seldom have only homogenous workload that just contains graph analysis. Often graph analysis is mixed with SQL, ML, and other analytics. 
	In order to support the heterogenous workload, developers have to move data around different systems, in the fragmented database world. By supporting multiple languages/APIs on the shared data, the converged database solution essentially allows users to view the data in the way that is needed! SQL, graph, and ML can work on the same data synergistically. There is also no data transfer or transformation needed. This is a huge saving. \rev{Even though native graph databases are highly optimized for graph workloads, but if we consider the performance of the end-to-end pipeline of a heterogenous workload, the converged graph databases may actually have an edge.}
	
	Moreover, some converged database solutions, such as IBM Db2 Graph, even allow graph queries to be performed on the original data in the operational databases. The extra advantage it brings is to have the graph query capability without disturbing the large number of existing relational applications and that transaction updates to the operational data can be visible to the graph queries in real time. 
	
	Other advantages of converged database solution come from the existing backend data store, for example transaction support, access control, compliance to audits and regulations, temporal support, scalability support, HA and DR support, etc. 

\rev{As discussed above, each type of graph solution has its own pros and cons. Choosing the right architecture depends highly on the actual application requirements, such as whether the workload is graph-only or heterogeneous, the latency and throughput requirement, the frequency of updates, the recency requirement of the results, etc.}

\section{Graph Benchmarks}\label{sec:benchmark}
Benchmarks are very important in evaluating different database systems. Since graph databases are a relative new area compared to relational databases, there is no standard benchmarks, like TPCC, TPCH, and TPCDS, yet for graph databases. \rev{There has been some community efforts in establishing graph benchmarks, such as the Linked Data Benchmark Council (LDBC)~\cite{ldbc} benchmarks, Linkbench from Facebook~\cite{linkbench}, Graph500~\cite{graph500}, HPC Scalable Graph Analysis Benchmark~\cite{hpcbench}, and Open Graph Benchmark~\cite{ogb} for graph ML specifically.} Notably, LDBC benchmarks are the most widely adopted and hence the closest to a standard. It is also the most comprehensive benchmark and currently has three benchmark suites. The LDBC Social Network Benchmark (LDBC-SNB) contains tests for interactive workload, which corresponds to graph OLTP workload, and business intelligent workload, which is more relational (aggregation and join heavy) queries on graph data and is still under development. The LDBC Graphalytics benchmark targets graph OLAP workload. And the LDBC Semantic Publishing Benchmark (SPB) is an RDF-based benchmark for semantic databases. Since LDBC is a non-profit organization with members from both industry and academia, graph benchmarking is another area where the graph research community has influenced the industry. Both TigerGraph and Neo4j has published white papers or blogs about their test results on LDBC-SNB benchmark~\cite{neo4jldbc, tigergraphldbc}, as well as the responses to each other's results~\cite{neo4jldbcresponse}.

\rev{This has been a lot of work in the research community that compares various graph databases~\cite{Dominguez-Sal, Jouili, Kolomicenko, microbench}. This line of research is highly useful, however, most of these efforts proposed their own benchmark frameworks, and the results from different studies often generated different conclusions. It will be more valuable if future studies adopt an existing widely used benchmark that the industry embraces, such as the LDBC benchmark.}



\section{Discussion}\label{sec:discussion}
	

\subsection{Graph Database Users}
There are different types of graph users from the most sophisticated to the most novice. The first category of graph users are the few companies such as LinkedIn or Facebook for whom \textit{graph is the business}! These users usually have an army of in-house engineers to build customized systems for their bread and butter~\cite{liquidblog1, tao}. They will not shop for a general graph database. The second category is the power users, where their core business has a strong dependence on graphs, \rev{such as companies specialized in fraud detection, anti-money laundering, security, and intelligence.} They most likely will adopt the best graph technologies in the market that fit their needs. The third category is a larger group of standard experienced users. Many of them were the traditional database customers who now build new applications enabled by graph technologies. \rev{Some example success stories of such customers were showcased in~\cite{tigergraphusecases}.} Then there is a large of number of novice users and potential users, who are interested in trying out graph technologies. The last three categories of users are the ones whom graph database vendors typically go after. And if they come from the traditional database side, converged database solutions may provide them with an easier entry point into graphs. To better serve the experienced graph users and convert novice or potential users, customers need education on what they can do with graph technologies and how to apply them. Pure-play companies like Neo4j and TigerGraph are particularly strong in customer education. They publish books, organize workshops, summits and conferences where they showcase the technologies as well as demonstrate use cases.

Another important point is that \textit{graph problems are greater than graphs}! Yes, customers need graph solutions, but not just graph solutions. They need an end-to-end solution which involves data ingestion, preprocessing, graph analytics, maybe also other types of analytics, and result rendering. When we think about the big picture, graph performance is not the only factor that customers consider when choosing graph solutions.

Finally, both graph OLTP and OLAP are important. Some customers primarily use one type vs the other, but others use both. For graph OLTP, due to the exploratory nature of the queries, graph visualization is a must-have. For graph OLAP, the set of built-in algorithms will be the winning factor that attracts customers. 

\subsection{In-House Graph Systems}

As mentioned before, big tech companies, like LinkedIn and Facebook, have developed their own specialized graph systems to serve their business needs. Although these in-house systems are not yet in the commercial space, these companies have a good track record of open-sourcing their in-house systems with great industrial impact. So, some of these graph systems are worth watching for.

LIquid~\cite{liquidblog1} is LinkedIn's in-house graph database for real-time querying of its economic graph. LIquid adopts a subject-predicate-object triple model to store edges similar to RDF, employs a declarative language based on Datalog, and achieves nanosecond-level query efficiency via dynamic query optimization on wait-free shared-memory index structures~\cite{liquid}. TAO~\cite{tao} is Facebook's geographically distributed graph system for serving efficient access to its social graph. It supports a property-graph-like model, provides simple APIs to access nodes and edges, and is built with an efficient caching layer on top of the MySQL storage layer.

\subsection{\rev{Research Graph Database Prototypes}}

\rev{Although a comprehensive survey of graph databases from academia is beyond the scope of this paper, it is still interesting to briefly discuss some recent related systems. Research graph systems are also divided into native and hybrid graph databases. Graphflow~\cite{graphflow} and AvantGraph \cite{avantgraph} are two examples of native graph systems. Both systems implemented the worst case optimal (wco) joins and factorization of intermediate results. Although not yet widely adopted in industry, these techniques have been proven to be promising in speeding up graph query processing and present great potential for future adoption.
GRFusion~\cite{grfusion} and GRainDB \cite{graindb} are two examples of hybrid graph databases. GRFusion~\cite{grfusion} extends VoltDB to define \emph{graph views} on relational tables, and to materialize graph structures in memory for the graph queries to execute on. This architecture resembles Oracle Spatial and Graph. GRainDB extends the internals of DuckDB to support predefined pointer-based joins, and plans to also incorporate wco joins and factorization into the query engine. Most commercial hybrid graph databases today have conservatively chosen to avoid modifying the core engines when supporting graph queries. GRainDB points out interesting directions on how graph-specific techniques can be incorporated into an existing database engine.}

\subsection{Opportunities and Directions}

There is a growing global market for graph databases. Right now, the graph-only vendors are leading in query performance and algorithm support. But there is still a lot of opportunities for new vendors or other existing vendors to catch up. Especially, when looking at end-to-end scenarios, data import and export may increasingly become a bottleneck for graph-only databases. Making data movement easier and faster will be a crucial investment for these pure-play vendors. Major cloud companies are also investing in the graph space. Their major advantage is that they own the \textit{whole stack} of data services, including operational data services (OLTP and NoSQL engines), which is a major source of graph data. It makes sense that most of these vendors adopt converged database solutions. Taking the advantage of the whole stack and focusing on end-to-end solutions will be a winning recipe for them. 

\subsection{Recommendation for Researchers}

The graph database research community has been focusing heavily on sophisticated algorithms and query performance so far. These are very important, but there are other equally important but also practical problems that the graph database industry cares about and needs more help on. 
\rev{Due to the connected nature of graphs, transaction support, especially in a distributed setting, is hard to achieve and even harder to perform efficiently in graph databases. Visualization is crucial for graph OLTP workloads, but laying out graphs in a way that users can clearly understand the relationship of entities in the graph is generally a hard problem. Compliance to regulations (e.g. GDPR) requires keeping track of versions of graph data and potentially supporting point-in-time queries, but this feature is unfortunately missing in most existing graph databases. Multi-tenancy and access control are not supported efficiently in existing graph databases (most just assume there is a single tenant for a graph database server), so helps are definitely needed on this front. Graph queries and analytics are seldom executed in isolation, so integrating them with non-graph workloads efficiently deserves a lot of attention. For native graph databases, optimizing data import and export will be crucial. }

\rev{In addition, existing graph research largely assumes read-only workloads. But real-life graphs do change over time. Future research would better support the dynamic nature of graphs. Some of the existing research also define their own graph models (e.g. simple graphs and attributed graphs) or propose new query languages. Focusing on widely adopted graph models and languages from industry might lead to more practical impact.}





	\scriptsize
	\bibliographystyle{abbrv}
	\bibliography{ref}  
	
\end{document}